# Highly efficient spin-orbit torque and switching of layered ferromagnet Fe$_3$GeTe$_2$

Mohammed Alghamdi[1*], Mark Lohmann[1*], Junxue Li[1], Palani R. Jothi[2], Qiming Shao[3], Mohammed Aldosary[1#], Tang Su[4,1], Boniface Fokwa[2] and Jing Shi[1]

1. Department of Physics and Astronomy, University of California, Riverside, California 92521, USA
2. Department of Chemistry, University of California, Riverside, California 92521, USA
3. Department of Electrical and Computer Engineering, University of California, Los Angeles, California 90095, USA
4. International Center for Quantum Materials, School of Physics, Peking University, Beijing 100871, P. R. China

**Among van der Waals (vdW) layered ferromagnets, Fe$_3$GeTe$_2$ (FGT) is an excellent candidate material to form FGT/heavy metal heterostructures for studying the effect of spin-orbit torques (SOT). Its metallicity, strong perpendicular magnetic anisotropy built in the single atomic layers, relatively high Curie temperature ($T_c$ ~ 225 K) and electrostatic gate tunability offer a tantalizing possibility of achieving the ultimate high SOT limit in monolayer all-vdW nanodevices. In this study, we fabricate heterostructures of FGT/Pt with 5 nm of Pt sputtered onto the atomically flat surface of ~ 15 – 23 nm exfoliated FGT flakes. The spin current generated in Pt exerts a damping-like SOT on FGT magnetization. At ~2.5x10$^{11}$ A/m$^2$ current density, SOT causes the FGT magnetization to switch, which is detected by the anomalous Hall effect of FGT. To quantify the SOT effect, we measure the second harmonic Hall responses as the applied magnetic field rotates the FGT magnetization in the plane. Our analysis shows that the SOT efficiency is comparable with that of the best heterostructures containing three-dimensional (3D) ferromagnetic metals and much larger than that of heterostructures containing 3D ferrimagnetic insulators. Such large efficiency is attributed to the atomically flat FGT/Pt interface, which demonstrates the great potential of exploiting vdW heterostructures for highly efficient spintronic nanodevices.**

*: These two authors contribute equally.

#: Present address: Department of Physics and Astronomy, King Saud University, Riyadh 11451, Saudi Arabia.

Fe$_3$GeTe$_2$ (FGT), a layered conducting ferromagnet, is an important member of the van der Waals (vdW) material family that has attracted a great deal of attention [1,2]. Similar to other known vdW ferromagnets such as Cr$_2$Ge$_2$Te$_6$ and CrI$_3$, FGT possesses the magnetic anisotropy perpendicular to the atomic layers which is retained down to monolayers. Different from the others, FGT stands out due to the following attractive properties. First, not only do FGT bulk crystals have the highest Curie temperature $T_c$ (~230 K), but monolayer FGT also has the highest $T_c$, (130 K), when compared to their vdW ferromagnetic counterparts [1,3,4]. Furthermore, the $T_c$ of thin FGT can be dramatically elevated to room temperature using electrostatic gating [2]. Second, few-layer thick FGT films have been successfully grown by molecular beam epitaxy [5], which makes ultimate wafer-scale monolayer all-vdW heterostructure fabrication possible. Third, while the other vdW magnets are semiconductors or insulators, FGT is a ferromagnetic metal which allows for studying its magnetism via magneto-transport measurements. In conventional devices using conducting ferromagnets with perpendicular magnetic anisotropy (PMA) such as CoFeB, spin-orbit torques (SOT) have been exploited for switching the magnetization [6]. SOT efficiency, the figure-of-merit for this application, contains both intrinsic properties such as the spin Hall angle of the heavy metals serving as the spin current source and extrinsic properties such as the transmission coefficient. The latter depends on the ferromagnet/heavy metal interface quality. Because of the vdW nature that provides atomically flat interface, FGT has the potential of having high SOT efficiency for switching its magnetization, especially in all-vdW heterostructures.

In this work, we investigate the SOT effects in FGT/Pt heterostructure devices containing thin exfoliated FGT and sputtered Pt. In such devices, the spin Hall effect in Pt produces a pure spin current which enters the FGT layer and exerts on it both field-like and damping-like torques [7]. Different from magnetic insulator devices in which the magnetization state is read out by the induced anomalous Hall effect (AHE) in Pt via proximity coupling [8,9], the large AHE response in FGT lends itself a sensitive detector of its own magnetization state. To quantify the effects of SOT, we carry out two types of measurements: pulsed current switching and second harmonic Hall measurements. From both measurements, we demonstrate that the SOT efficiency in FGT/Pt is significantly larger than that in devices containing conventional three-dimensional (3D) magnets. In addition, we have observed SOT-induced switching of FGT magnetization with high switching efficiency.

Fe$_3$GeTe$_2$ crystals were grown by solid-state reaction of the elements at 800 $^o$C within 5 days. After mixing the elements Fe, Ge and Te in their stoichiometric molar ratio, the mixture was pressed into a pellet, sealed in a quartz glass ampoule under vacuum and loaded into the furnace for reaction. Figure 1 (a) shows the X-ray diffraction (XRD) pattern of a bulk FGT single crystal which agrees with the literature [10-14]. The XRD pattern contains only the (0 0 2n) Bragg peaks (n=1, 2, 3, 4, 5, 6), indicating that the exposed surface is the *ab*-plane of the FGT crystal. Indexation of the peaks led to the *c* lattice parameter of 16.376 Å, which is consistent with previously reported value [10]. To characterize the magnetic properties of FGT, we have carried out AHE measurements. The fabrication consists of the following steps. We start with FGT crystals. After exfoliation, we locate a desired flake and perform electron beam lithography (EBL) and lift-off to fabricate Pt (30 nm) contacts to the chosen FGT flake. The process for the FGT-only devices is similar to what will be illustrated in Fig. 2 for FGT/Pt heterostructure devices except that there are fewer steps here.

The hysteresis loops of the anomalous Hall resistivity $\rho_H$ for a FGT device with thickness of 53 nm are displayed in Figs. 1(b) and (c) for different temperatures ranging from 2 K to 230 K (device image is shown in the inset of Fig. 1(d)). Below 180 K, the $\rho_H$ loops are squared with monotonically increasing coercive field $H_c$ as the temperature is decreased. $H_c$ reaches ~ 7.5 kOe at 2 K, indicating very strong PMA. At 180 K, where we perform all SOT measurements to be presented later, $H_c$ is ~0.65 kOe. In hard-axis Hall measurements, we find the saturation field, denoted as $H_k$, to be ~30 kOe, which is 46 times larger than $H_c$. Above 180 K, the $\rho_H$ loops deviate from the squared shape, collapse at ~210 K, and finally disappear at ~230 K. In the meantime, the magnitude of $\rho_H$ loops, i.e., the height between the two saturated values, decreases as the temperature is raised, and vanishes at the Curie temperature $T_c$ as illustrated in Fig. 1(d). $T_c$ of this FGT device is found to be ~225 K. A more accurate determination of $T_c$ from the Arrott plot gives $T_c$ =224.5 K for the same device (See Supplementary Fig. 1). The overall temperature dependence of $\rho_H$ resembles but slightly steeper than the mean-field magnetization of FGT (see Supplementary Fig. 3). We note that the low-temperature $M_s$ value ranges from 285 to 393 emu/cm$^3$ [11,15-19]. Since most of our SOT experiments are carried out at 180 K, we take $M_s$=170 emu/cm$^3$ at 180 K from ref. 11, which is the lower-bound $M_s$ value for FGT. Using this $M_s$ value and the measured anisotropy field $H_k$, we obtain the minimum uniaxial PMA energy of $1.1 \times 10^7$ erg/cm$^3$ at 180 K, which is nearly two orders of magnitude greater than that of CGT of $1.4 \times 10^5$ erg/cm$^3$ at ~ 4 K [20].

To fabricate FGT/Pt bilayer devices for the SOT study, we adopt the fabrication processes as represented in Fig. 2. FGT flakes are first exfoliated from a small crystal shown in Fig. 2(a) and placed on a Si/SiO$_2$ wafer. As schematically shown from Fig. 2(b) to 2(e), a suitable flake is chosen (Fig. 2(b)) and covered with a 5 nm layer of Pt (Fig. 2(c)) by sputtering. Cr (5 nm)/Au (85 nm) electrodes are formed by EBL, e-beam evaporation, and liff-off (Fig. 2(d)). The continuous Pt film covering the flake is etched by inductively coupled plasma to form isolated Cr/Au electrodes (Fig. 2 (d)). The scanning electron micrograph of a final device is shown in Fig. 2(f). Atomic force microscopy (AFM) imaging of both FGT and FGT/Pt (See Supplementary Fig. 2) indicates atomic level flatness with the root-mean-square roughness of 0.2 nm, which is smaller than the atomic step height of FGT (0.8 nm) [2].

Fig. 3 (a) is the schematic illustration of our FGT/Pt device for the SOT study. When a charge current passes in both Pt and FGT layers, the former generates SOTs to act on the magnetization of the latter. In the pulsed current switching experiments, we pass current pulses increasing in amplitude and interrogate the FGT magnetization state by measuring the AHE resistivity, $\rho_H$, after each pulse through a small constant current bias. As the current reaches a threshold, the magnetization state of FGT switches and produces a sign reversal of $\rho_H$. We measure the critical currents for different in-plane fields. To more accurately determine the current flowing in Pt which is responsible for the SOT acting on FGT, we use a parallel resistor model with resistivities measured separately for 5 nm Pt on SiO$_2$ and 53 nm FGT flake (See Supplementary Fig. 3).

Before turning on sizable SOT, we first prepare the initial state of the FGT magnetization by applying an in-plane field $H_x$. Fig. 3(b) is the AHE response of the FGT(15 nm)/Pt(5 nm) device to an $H_x$ sweep measured at 180 K with a 50 µA current, which produces negligible SOT. This is a typical hard-axis hysteresis loop for materials with PMA. The easy-axis $\rho_H$ hysteresis loops are very similar to those of the FGT-only device shown in Fig. 1 except that the presence of the Pt layer provides a shunting channel which reduces the $\rho_H$ magnitude. At $H_x$=0, $\rho_H$ retains the full saturation value of FGT/Pt for the easy-axis field sweeps, indicating that the initial magnetization is perpendicular to the *ab*-plane of the FGT. With a sufficiently strong $H_x$ field, the magnetization is aligned to $H_x$ which results in a vanishing $\rho_H$. This saturation field $H_k$ is related to the strength of PMA field $H_u$ by $H_k = H_u - 4\pi M_s$. At an intermediate in-plane field $H_x = \pm 10$ kOe, the perpendicular component of the magnetization is reversed, which is caused by the incidental z-component of the applied magnetic field due to the misalignment of the applied field with the *ab*-plane. In our pulsed current switching experiments, we set the $H_x$ field bias below this threshold and then apply current pulses to generate additional SOT fields to induce switching. Clearly, the effective field from the damping-like SOT, i.e., $H_{DL} \sim \boldsymbol{\sigma} \times \boldsymbol{m}$, is responsible for the switching, here $\boldsymbol{\sigma}$ being the spin polarization direction of the spin current and $\boldsymbol{m}$ being the unit vector of the FGT magnetization. The critical current density $J_c$ required to switch the magnetization depends on the magnitude of $H_x$. The full $H_x$-current switching phase diagram is shown in Figs. 3(e) and 3(f). Figs. 3(c) and 3(d) are the line cuts for three selected $H_x$ fields: ±3, ±6, and ±9 kOe. At $H_x$ = -9 kOe, switching occurs at $J_c \sim 1.5 \times 10^{11}$ A/m². Here the $J_c$ value is the critical current density in Pt. If the strength of $H_x$ is decreased to 3 kOe in the negative direction, $J_c$ increases to ~ $2.0 \times 10^{11}$ A/m². We extrapolate $J_c$ linearly to $H_x$=0 along the line shown in Fig. 3 (e) and find $J_c(H_x=0) = 2.5 \times 10^{11}$ A/m². A similar $J_c$ value is found for the positive $H_x$ side, by performing the same extrapolation in Fig. 3(f). To compare the effectiveness of the SOT in switching, we calculate the switching efficiency parameter $\eta$ using $\eta = \frac{2eM_s t H_c}{\hbar J_c(H_x=0)}$ [8], representing the ability of switching the magnetization with SOT. $H_c$ is ~0.65 kOe for FGT at 180 K, much smaller than $H_k$ (30 kOe), indicating that switching is by domain nucleation and domain wall depinning. If again taking the lower-bound value for $M_s$ of 170 emu/cm³ for our FGT/Pt device, we obtain a minimum $\eta$ value of 1.66. $\eta$ can be as high as 2.2 if $M_s$ is taken to be 225 emu/cm³ at 180 K [15]. These $\eta$ values are higher than those reported in $Tm_3Fe_5O_{12}$/W (0.95) [8] and $Tm_3Fe_5O_{12}$/Pt (0.014) [9] and suggest highly efficient SOT switching of FGT magnetization via local domain wall depinning.

To further quantify SOT, we perform second-harmonic (2ω) Hall measurements on FGT/Pt devices with the measurement geometry shown in Fig. 4(a). More details and application of the method were described in refs. [21,22]. We measure the 2ω responses in the Hall resistance, here ω being the frequency of the AC current passing through the device. The 2ω signal is present only if there is a SOT acting on the magnetization. This harmonic signal is recorded as a function of a rotating in-plane magnetic field. We rotate the magnetization with an in-plane magnetic field of fixed magnitudes that are higher than $H_k$ and measure the second harmonic Hall signal $R_H^{2\omega}$.

As indicated in Eq. 1 below, $R_H^{2\omega}$ consists of both $cos\varphi$ and $cos(3\varphi)$ terms, here $\varphi$ being the azimuthal angle between the magnetic field and current direction,

$$R_H^{2\omega} = \left[R_{DL}^{2W} + R_{TH}^{2W} + \frac{R_{FL}^{2W}}{2}\right] * cos\varphi + \frac{R_{FL}^{2W}}{2} cos(3\varphi). \qquad (1)$$

In Eq. 1, the $cos\varphi$ term contains the damping-like SOT contribution $R_{DL}^{2\omega}$ via AHE, thermoelectric contribution $R_{TH}^{2\omega}$ via anomalous Nernst effect, and the field-like SOT contribution $R_{FL}^{2\omega}$ via the planar Hall effect. The $cos(3\varphi)$ term contains only the field-like SOT contribution $R_{FL}^{2\omega}$. Fig. 4(b) and 4(c) display the total $R_H^{2\omega}$ signals from FGT(23 nm)/Pt(5 nm) device for different magnetic fields with the AC current amplitudes of 2.2 mA and 2.4 mA, respectively. These results can be fitted very well by the $cos\varphi$-function only, indicating the negligible effect from the field-like SOT, which is commonly true for ferromagnetic metal/heavy metal heterostructure. Further analysis of the external field strength dependence allows us to separate the damping-like SOT effect from the thermal effect, as shown in Fig. 4(d), which yields an effective SOT field $H_{DL}$ for each current. Using the smallest $M_s$ value of 170 emu/cm$^3$ for FGT at 180 K, we calculate the lower-bound damping-like torque efficiency $\xi_{DL}$ in FGT/Pt bilayer and obtain $\xi_{DL}$=0.11±0.01 for 2.2 mA and $\xi_{DL}$=0.14±0.01 for 2.4 mA. In our $\xi_{DL}$ calculations, we only use the current in Pt based on the parallel resistor model; therefore, it should be valid to compare this $\xi_{DL}$ for FGT/Pt with the available $\xi_{DL}$ values for both ferrimagnetic insulator/heavy metal and ferromagnetic metal/heavy metal heterostructures. We note that even the minimum $\xi_{DL}$ value for FGT/Pt is significantly larger than $\xi_{DL}$ in Tm$_3$Fe$_5$O$_{12}$/Pt (0.058 in [9] and 0.015-0.02 in [23]). Interestingly, our minimum $\xi_{DL}$ compares very well with the highest value of ≈0.15 for CoFeB/Pt in literature [24].

Both the switching efficiency $\eta$ and SOT efficiency $\xi_{DL}$ in FGT/Pt are higher than or comparable with those in conventional SOT devices fabricated with 3D magnetic materials. It is worth pointing out that the single-domain requirement for Eq. 1 is fulfilled in the second harmonic Hall measurements, so that $\xi_{DL}$ extracted from our experiments is reliable. By using the minimum $M_s$, this $\xi_{DL}$ represents the lower bound value for SOT efficiency. The reason for this very high SOT efficiency in FGT/Pt is currently not completely understood. Here we believe that the excellent interface resulting from atomically flat FGT surface plays an important role; therefore, the high SOT efficiency may be common to heterostructures fabricated with other vdW ferromagnets.

In summary, using both pulsed current switching and harmonic Hall measurements, we have demonstrated highly efficient SOT effects and magnetization switching in heterostructures containing a few-layer vdW ferromagnet and Pt. Since the atomic flatness of the vdW ferromagnets is an inherent property of the materials, it is expected that the high-quality interface can be retained even down to monolayers. Due to the strong PMA, switching of monolayer FGT can be potentially achieved with a much lower critical current density, which leads to much more efficient spintronic nanodevices.

## Methods

### Device fabrication

For the FGT device, the flake is exfoliated onto a Si/SiO$_2$ substrate followed directly by spin coating 200 nm of PMMA and baking on a hotplate in air at 120 ºC for 3 minutes. This low temperature helps protect the FGT flake from degradation and oxidation. Electrode patterns are then formed by EBL followed by sputtering a 30 nm of Pt. Before deposition of the electrodes, the contact region is plasma cleaned in the sputtering chamber with 15W Ar plasma at a pressure of 30 mTorr for 30 sec. Directly after liftoff, the device is mounted and loaded into an evacuated cryostat where the transport measurements are performed.

For the FGT/Pt devices, the flakes are exfoliated onto a Si/SiO$_2$ substrate and instantly transferred into the loadlock of our sputtering system which is evacuated to a base pressure of $10^{-7}$ Torr. Once the base pressure is reached, the entire substrate is plasma cleaned with 15 W Ar plasma at a pressure of 30 mTorr for 30 sec. Then a 5 nm layer of Pt is sputtered forming a continuous Pt film on the substrate. Once removed from the sputtering chamber, an optimal FGT/Pt flake is chosen by optical microscope and then EBL is performed to define an electrode pattern followed by immediate deposition of Cr(5 nm)/Au(85 nm) by electron beam evaporation. One last EBL step is then performed to define a mask to etch the FGT/Pt flake into the Hall geometry and remove all Pt connections between the electrodes. Inductively coupled plasma etching with Ar is then performed on the device and the completed device is placed into an acetone bath to remove the PMMA mask.

### Electrical transport measurements

All transport measurements for the FGT and FGT/Pt devices are performed in the Physical Properties Measurement System by Quantum Design in a temperature range of 300 K to 2 K. For the FGT device we kept a fixed current of 50 µA in the flake with a Keithley 2400 source meter which also monitored the two-terminal resistance. To monitor the longitudinal and Hall resistances two Keithley 2182A nanovoltmeters were used. For the DC switching measurements in the FGT/PT heterostructures, a similar setup was used to monitor the response of the Hall and longitudinal resistances while a Keithley 6221 AC current source was used to pulse a square 0.5 micro second DC current through the device. For the 2ω Hall measurement, we fixed a constant

AC current at a frequency of 13.113 Hz in the device with the Keithley 6221 AC current source. The 1ω and 2ω Hall responses were monitored with two Stanford Research SR830 AC lock-ins.

**Acknowledgements**

The authors would like to acknowledge support by DOE BES Award No. DE-FG02-07ER46351 for device nanofabrication, transport measurements, and data analyses, and by NSF-ECCS under Awards No. 1202559 NSF-ECCS and No. 1610447 for the construction of the pickup-transfer optical microscope and device characterization.
.

**Author contributions**

J.S. designed and supervised the project. M. Alghamdi and M.L. fabricated nanodevices and performed switching and harmonic measurements. J.X.L., Q.M.S., M. Aldosary, and T.S. provided assistance in performing the experiments and data analysis. P.J. grew the FGT crystals and performed the XRD experiments under the supervision of B.F. M. Alghamdi., M.L, J.X.L, and J.S. wrote the manuscript and all the authors contributed to the final version of manuscript.

**Additional information**

Supplementary information is available.

**Competing interests**

The authors declare no competing interests.

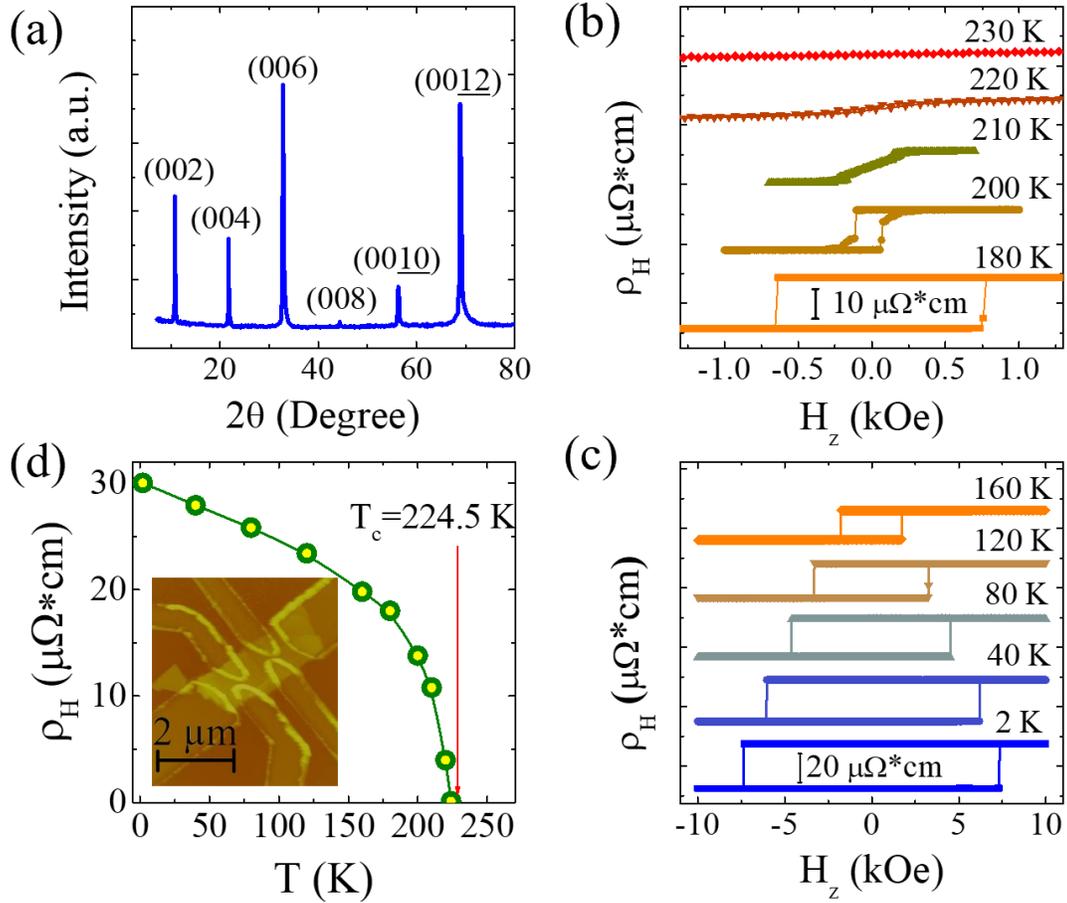

**Figure 1. Characterization of FGT bulk crystal and flakes**. (a) X-ray diffraction pattern for bulk FGT single crystal's ab-plane. (b,c) Hall resistivity as a function of applied field for a 53 nm thick flake of FGT at selected temperatures from 2 K – 230 K. (d) Measured Hall resistivity as a function of temperature for the same FGT device with $T_c$ determined by the Arrott plot (See supplementary Fig. 1) labeled. Inset shows the AFM topographic image for our FGT device.

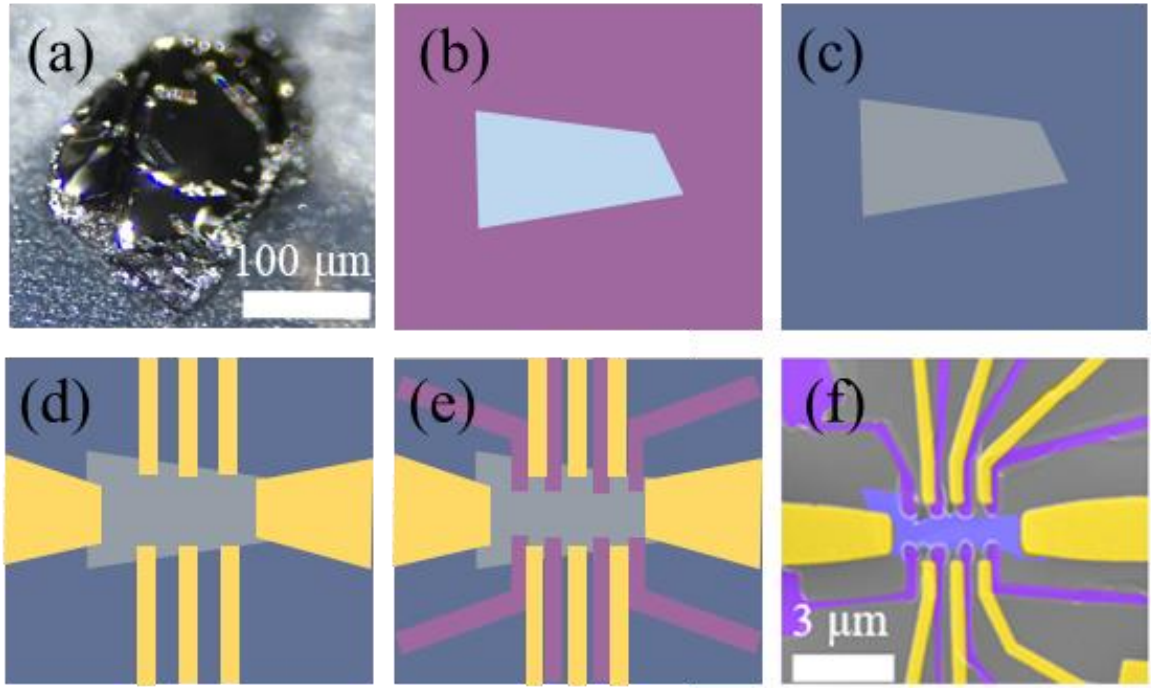

**Figure 2. Fabrication process for FGT/Pt hybrid devices**. (a) Optical microscopic image of a bulk FGT piece on scotch tape. (b-e) Illustration of the fabrication process. First, FGT flakes are exfoliated onto 300 nm thick $SiO_2$ substrates and a suitable flake is located, (b), followed by sputtering 5 nm of Pt, (c), then electrodes are placed on the flake, (d), and lastly the Pt and FGT is etched in order to define the Hall geometry and to remove any Pt connections between the electrodes, (e). (f) False-colored SEM image of the FGT(15 nm)/Pt(5 nm) device used for SOT-induced magnetization switching.

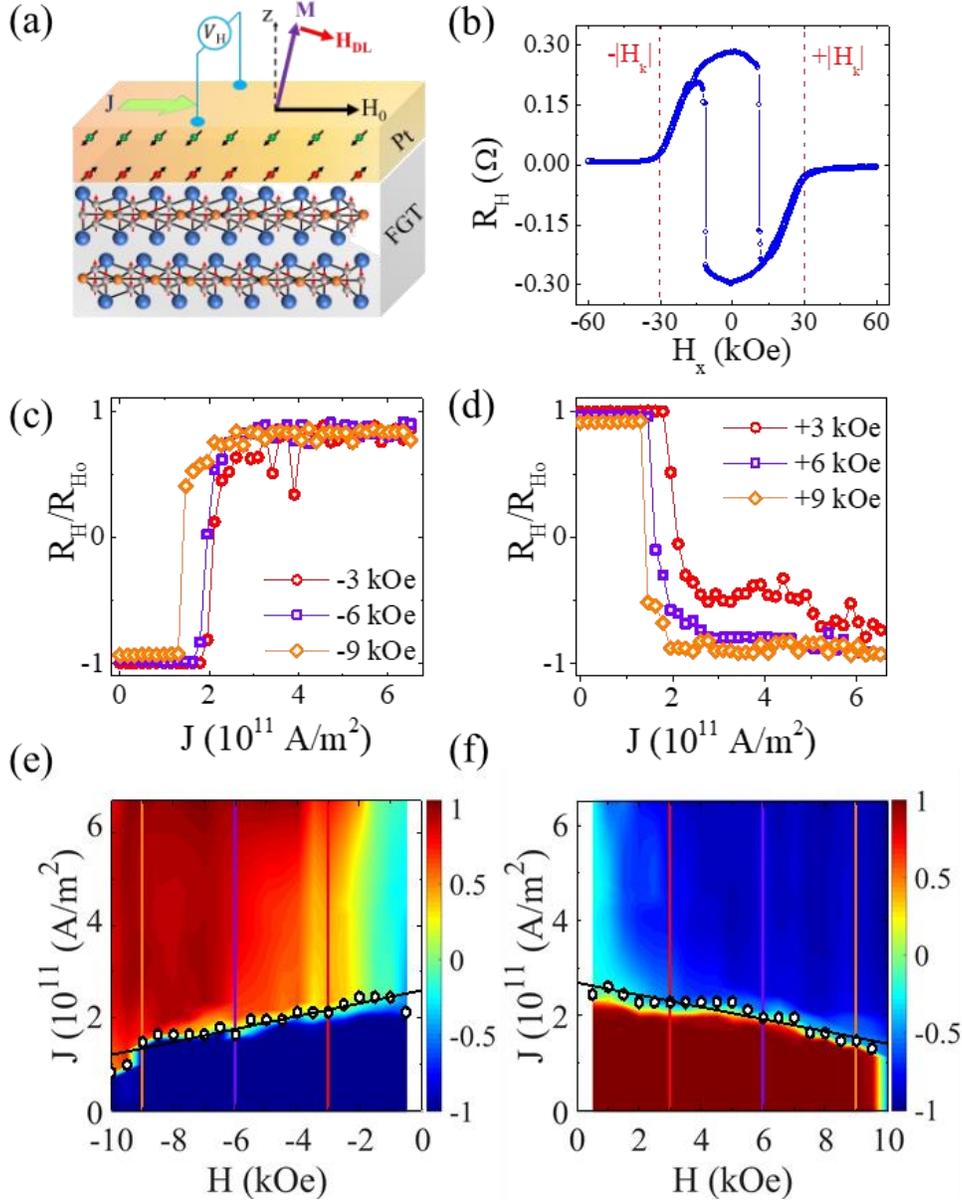

**Figure 3. Current-induced magnetization switching of FGT**. (a) Schematic illustration of the effective field responsible for switching the magnetic state of FGT in our FGT/Pt hybrid devices. $J$ is the injected current density, $H_x$ is the applied in-plane field, $H_{DL}$ is the effective field from damping-like SOT, and M is FGT's magnetization. (b) Hall resistance for an applied in-plane magnetic field at 180 K for our FGT(15 nm)/Pt(5 nm) device (showing in Figure 2. (f)) with anisotropy field $H_k$ labeled on the graph. (c-f) Effective switching current as a function of applied in-plane negative, (e), and positive, (f), bias field. The color scale represents the switching resistance as a percentage of the absolute value of the anomalous Hall resistance at zero current $R_{H0}$. (c,d) correspond to the line cuts in (e,f).

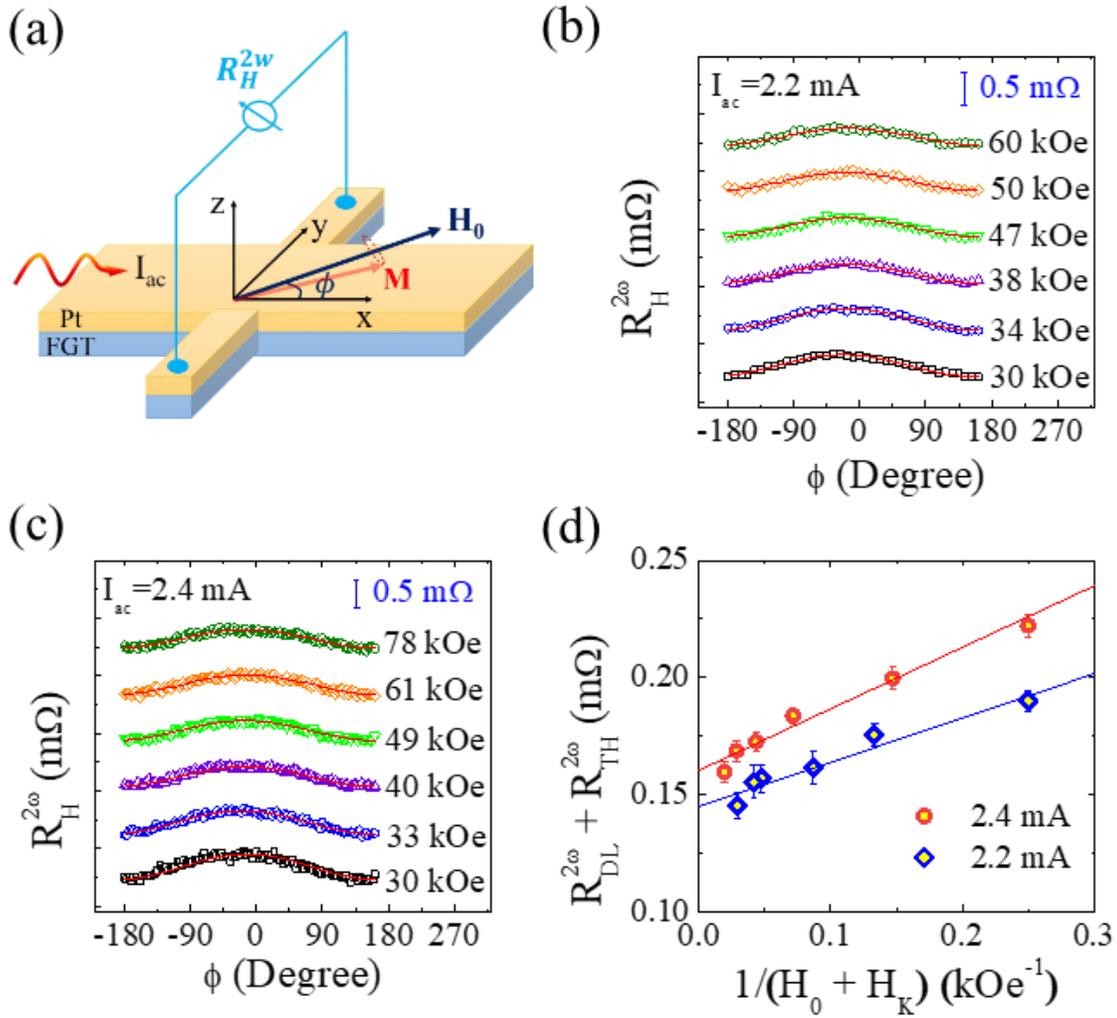

**Figure 4. Determination of SOT efficiency from second harmonic Hall measurements.** (a) Schematic illustration of $2\omega$ Hall measurement in FGT/Pt device. $I_{AC}$ represents the injected AC current amplitude, $H_0$ and $M$ the applied in-plane field and magnetic moment respectively, and $\varphi$ the azimuthal angle. (b,c) $2\omega$ Hall resistance of a FGT(23 nm)/Pt(5 nm) device as a function of azimuthal angle for $I=2.2$ mA, (b), and $I=2.4$ mA, (c). The symbols represent the raw data and red solid lines represent the fit to the theoretical model using Eq. 1. (d) Amplitude of the $\cos(\varphi)$ coefficient from the fitting as a function of $1/(H+H_K)$. The slope is used to determine the SOT efficiency of the system.